\documentclass[prl,amsmath,amssymb,superscriptaddress]{revtex4}

\usepackage{graphicx}
\usepackage{dcolumn}
\usepackage{bm}
\usepackage{hyperref}
\usepackage{multirow}

\newcommand*\us{$\mu$s}
\newcommand*\Mn{Mn$^{2^+}$}
\newcommand*\MnCl{MnCl$_2$}

\newcommand*\tone{\ensuremath{T_1}}
\newcommand*\ttwo{\ensuremath{T_2}}
\newcommand*\tstar{\ensuremath{T_2^*}}
\newcommand*\fe{\ensuremath{f_e}}
\newcommand*\Bsd[1]{\ensuremath{B_{\mathrm{#1}}}}

\newcommand*\GL[1]{\ensuremath{\Gamma_{\mathrm{#1}}}}
\newcommand*\GUL[2]{\ensuremath{\Gamma^{\mathrm{#1}}_{\mathrm{#2}}}}
\newcommand*\gNV{\ensuremath{\gamma_{_{\mathrm{NV}}}}}

\newcommand*\colint{green}
\newcommand*\colext{grey}
\newcommand*\coltot{red}

\newcommand*\colmn{red}
\newcommand*\colair{green}
\newcommand*\colwater{blue}
\newcommand*\colacc{black}
\newcommand*\colhcl{orange}
\begin{document}

\title{Ambient Nanoscale Sensing with Single Spins Using Quantum Decoherence}

\author{L. P. McGuinness}
\altaffiliation{These authors contributed equally}
\affiliation{Centre for Quantum Computation and Communication Technology, School of Physics, University of Melbourne, Victoria 3010, Australia}
\affiliation{Institut f\"{u}r Quantenoptik, Universit\"{a}t Ulm, 89073 Ulm, Germany}
\author{L. T. Hall}
\altaffiliation{These authors contributed equally}
\affiliation{Centre for Quantum Computation and Communication Technology, School of Physics, University of Melbourne, Victoria 3010, Australia}
\author{A. Stacey}
\affiliation{School of Physics, University of Melbourne, Victoria 3010, Australia}
\affiliation{Element Six Ltd, King's Ride Park, Ascot SL5 8BP, UK}
\author{D. A. Simpson}
\affiliation{Centre for Quantum Computation and Communication Technology, School of Physics, University of Melbourne, Victoria 3010, Australia}
\author{C. D. Hill}
\affiliation{Centre for Quantum Computation and Communication Technology, School of Physics, University of Melbourne, Victoria 3010, Australia}
\author{J. H. Cole}
\affiliation{Chemical and Quantum Physics, School of Applied Sciences, RMIT University, Melbourne 3001, Australia.}
\affiliation{Institut f\"{u}r Theoretische Festk\"{o}rperphysik und DFG-Center for Functional Nanostructures (CFN), Universit\"{a}t Karlsruhe, D-76128 Karlsruhe, Germany}
\author{K. Ganesan}
\affiliation{School of Physics, University of Melbourne, Victoria 3010, Australia}
\author{B. C. Gibson}
\affiliation{School of Physics, University of Melbourne, Victoria 3010, Australia}
\author{S. Prawer}
\affiliation{School of Physics, University of Melbourne, Victoria 3010, Australia}
\author{P. Mulvaney}
\affiliation{School of Chemistry, Bio21 Institute, University of Melbourne, Parkville Victoria 3010, Australia}
\author{F. Jelezko}
\affiliation{Institut f\"{u}r Quantenoptik, Universit\"{a}t Ulm, 89073 Ulm, Germany}
\author{J. Wrachtrup}
\affiliation{3. Physikalisches Institut, Research Center SCOPE, and MPI for Solid State Research, University of Stuttgart, Pfaffenwaldring 57, 70569 Stuttgart, Germany}
\author{R. E. Scholten}
\affiliation{Centre for Coherent X-ray Science, School of Physics, University of Melbourne, Victoria 3010, Australia.}
\author{L. C. L. Hollenberg}
\affiliation{Centre for Quantum Computation and Communication Technology, School of Physics, University of Melbourne, Victoria 3010, Australia}
\email[]{lloydch@unimelb.edu.au}

\date{\today}

\maketitle

Magnetic resonance detection is one of the most important tools used in life-sciences today. However, as the technique detects the magnetization of large ensembles of spins it is fundamentally limited in spatial resolution to mesoscopic scales. Here we detect the natural fluctuations of nanoscale spin ensembles at ambient temperatures by measuring the decoherence rate of a single quantum spin in response to introduced extrinsic target spins. In our experiments 45 nm nanodiamonds with single nitrogen-vacancy (NV) spins were immersed in solution containing spin 5/2 Mn$^{2+}$ ions and the NV decoherence rate measured though optically detected magnetic resonance. The presence of both freely moving and accreted Mn spins in solution were detected via significant changes in measured NV decoherence rates. Analysis of the data using a quantum cluster expansion treatment of the NV-target system found the measurements to be consistent with the detection of ~2,500 motionally diffusing Mn spins over an effective volume of (16 nm)$^3$ in 4.2 s, representing a reduction in target ensemble size and acquisition time of several orders of magnitude over state-of-the-art electron spin resonance detection. These measurements provide the basis for the detection of nanoscale magnetic field fluctuations with unprecedented sensitivity and resolution in ambient conditions.\\

The generation and dynamics of fluctuating nanoscale magnetic fields is central to many important processes in biology, from ion-channel function and free-radical formation in the intra-cellular medium to neurons firing in the brain. Yet, there is a critical lack of techniques capable of detecting the small numbers of electronic or nuclear spins that are at the heart of these processes under physiological conditions. Under high vacuum and sub-Kelvin operating conditions, magnetic resonance force microscopy can detect single electron \cite{rugar04} and as few as 100 nuclear spins in the solid-state \cite{mamin07} with 3D imaging capability \cite{degen09} setting the absolute benchmark for the best spin detection and imaging sensitivity. However, achieving magnetic detection of small ensembles of spins in solution under ambient conditions is a significant challenge. Magnetic resonance imaging techniques have made tremendous progress with imaging of micron sized iron oxide particles in single cells \cite{shapiro04}, although detection is still limited to mesoscopic scales $\sim 10$\,$\mu$m because of the relatively large spin ensembles of order $10^7$ required to obtain sufficient signal \cite{blank04}.

\begin{figure*}
\centering
\includegraphics[width=\textwidth]{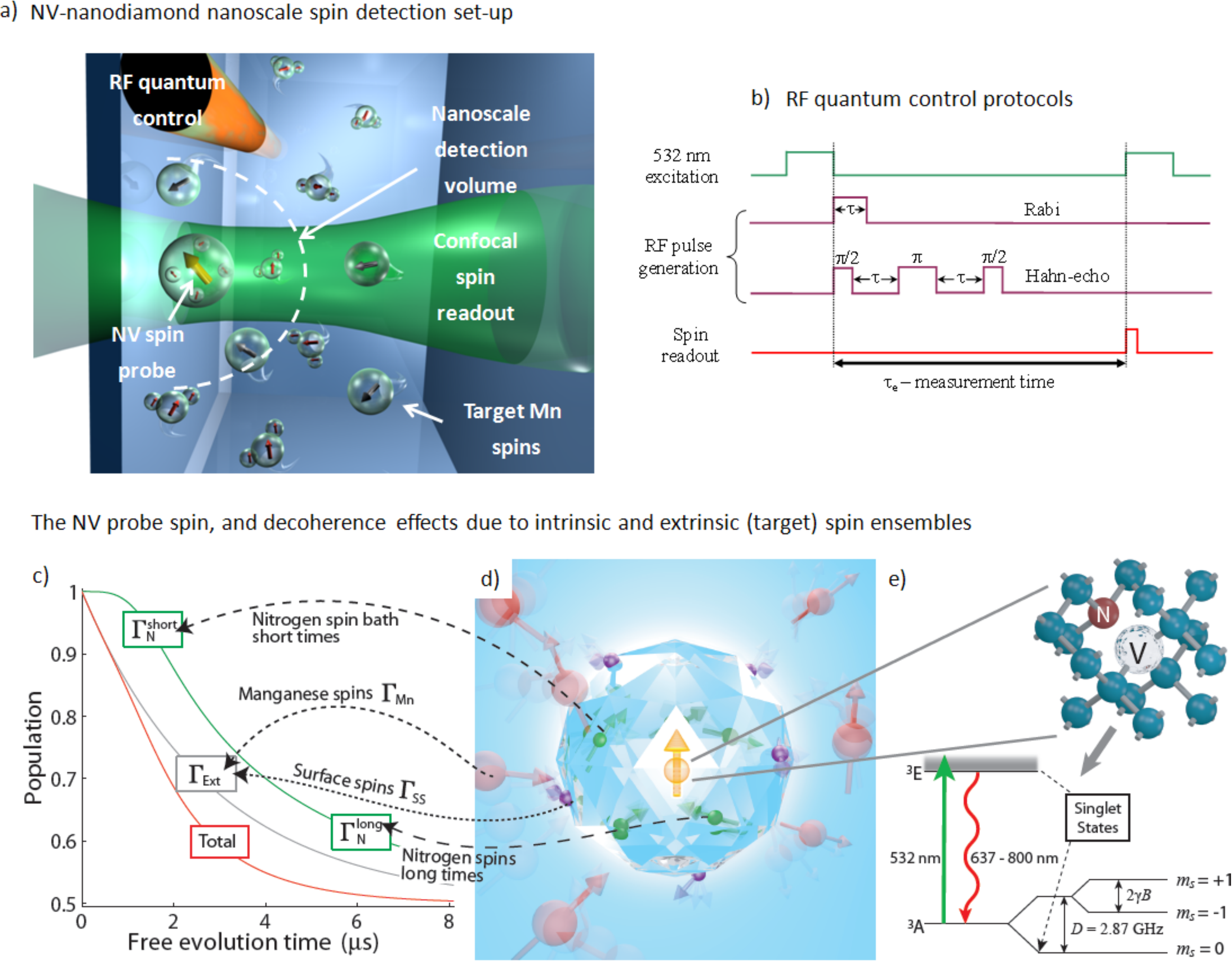}
\caption{Overview of nanoscale decoherence sensing with NV spins. (a) Experimental set-up: NV nanodiamonds are introduced into a spin-rich MnCl$_2$ solution (spin target). The NV spin is controlled by an RF microwave line and read-out optically using a confocal system. (b) Quantum control protocols to measure the decoherence of the NV spin in response to the local magnetic field fluctuations. (c) Environmental spin baths can be detected by their effect on the NV coherence. The decoherence contribution from internal spins (\colint), external spins (\colext), and total effect (\coltot) on NV coherence (\ttwo) time measured with spin-echo. (d) Magnetic sources local to the NV centre include diffusing \Mn{} spins (red), surface spins (blue) and internal nitrogen spins (green). (e) Physical and electronic energy structure of the NV centre in diamond (hyperfine structure not shown).}
\label{fig:principle}
\end{figure*}

Here we report nanoscale detection of order $10^3$ electronic spins under ambient conditions by measuring the changes in decoherence rate (\ttwo) of a single spin probe in response to extrinsic target spins in its local environment. In our experiments we introduced \Mn{} spins in solution around nanodiamonds containing a single-spin NV defect and detected diffusion and accretion of the target \Mn{} spins by measuring changes in the NV decoherence time (Fig.\,\ref{fig:principle}). The technique of decoherence sensing \cite{cole09} exploits the sensitivity of individual quantum spin systems to minute variations in the magnetic environment, and even permits detection of fields with zero mean, by measuring fluctuations in field amplitude \cite{hall09}. In contrast to conventional MR imaging, where variations in spin relaxation (\tone) and/or dephasing (\ttwo, \tstar) processes between different macroscopic regions form the basis for contrast, detection through measuring the decoherence of a single spin probe is inherently nanoscopic due to the $1/r^3$ fall-off of the dipole interaction. A useful comparison is to conventional electron spin resonance (ESR) where the local environment of ensembles of spin moieties, directly measured by magnetic resonance, are deduced in an average sense over the typically large spin ensembles required for detection, whereas our technique uses magnetic resonance on a single spin probe qubit to indirectly detect the nanoscale spin ensemble surrounding the probe. As advances in nanotechnology allow routine manipulation of single atomic systems, the interaction between single qubits and local spin baths have been investigated for both nuclear \cite{kloppens08, cywinski09PRL, bluhm11, childress06, bala09, mizuochi09} and electronic \cite{hanson08Science,delange10,morello10} spin baths in various semiconductors. However, these studies were restricted to spins already present in the host medium, and detection of general environments of interest has not been achieved. More recently, coherent interaction between a single NV and a spin on the diamond surface has been reported \cite{grotz11}. This is an important step towards the ultimate goal of the detection of spins and associated processes extrinsic to the quantum probe medium.

We present the results of three experiments. First we demonstrate decoherence sensing with single spins by sequentially characterising changes in the local NV magnetic environment for nanodiamonds in air, water (milliQ), and high-spin \MnCl{} using a lab-built confocal microscope with microwave control. We then confirm these results by restoring the coherence of two NV sensors by washing with spin-free acid. Finally, we use decoherence measurements to detect accretion of \Mn{} ions onto the nanodiamond surface. A full quantum cluster expansion treatment of the NV interaction with the intrinsic and external spin baths is used to analyse the measurements. The results show that the NV-nanodiamonds used here are detecting the randomly fluctuating magnetic fields produced by nanoscale spin ensembles of order 2,500 spins at ambient temperature with single-time-point detection times of a few seconds. This is an improvement of several orders of magnitude in resolution over state-of-the-art MR detection of electronic spins \cite{blank04}. Just as MR detection and imaging has made critical contributions to biological imaging down to the mesoscale, the capabilities of decoherence based magnetic sensing and imaging using atomic sized NV quantum probes may offer important new opportunities for detection in nanoscale biology.

\begin{figure*}
\centering
\includegraphics[width=\textwidth]{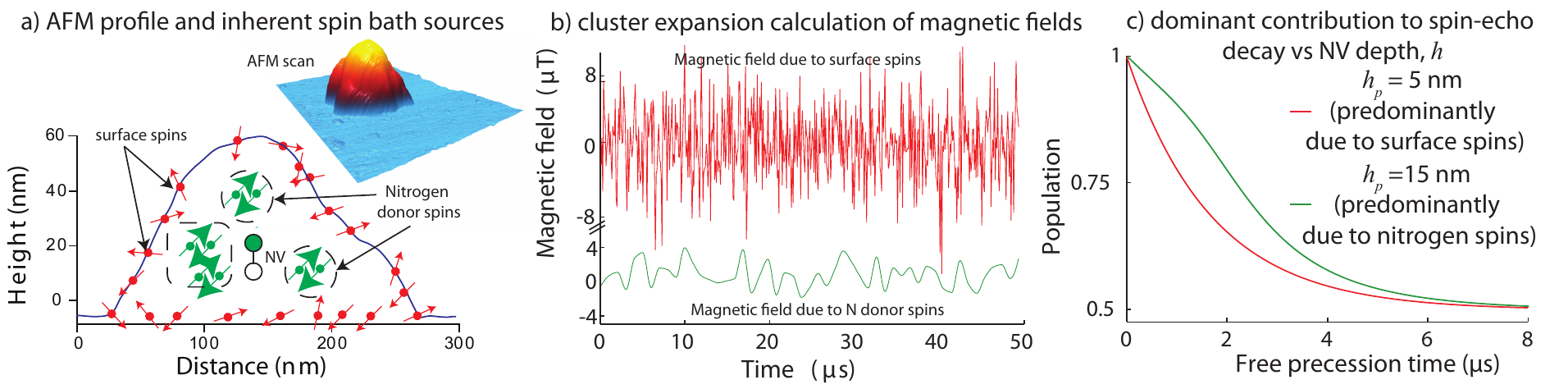}
\caption{Topology of nanodiamond sensors, inherent spin baths, and cluster expansion simulations of their effect on NV coherence. (a) Atomic Force Microscope (AFM) profile (blue curve and insert) of an agglomeration resulting from spin-coating nanodiamond (45nm median diameter) onto a glass substrate. Overlaid with the AFM data is a schematic depicting the intrinsic crystal spin-baths comprising strongly interacting clusters of internal nitrogen (green) and surface spins (red). (b) Cluster expansion simulation of the resulting typical magnetic fields felt by an NV centre (depth $h=$ 5 nm in this case) due to the various intrinsic spin-bath sources. Surface spins are distributed with a much greater density than the internal spins, and the fluctuation frequency of the surface spins is much faster than the nitrogen spin bath. In these simulations we have assumed a mean fluctuation rate of 100 MHz and a spin density of 1 nm$^{-2}$ for the surface spins. (c) NV spin-echo envelopes corresponding to the cluster state simulations in (b) as a function of depth in the nanocrystal. NV centres closer to the surface have a shorter coherence time, due to proximity to surface spin bath.}
\label{fig:theory}
\end{figure*}

\section{\label{theory}Decoherence sensing with nv centres in diamond}
The NV centre is an atomic sized defect occurring naturally in diamond, comprised of a substitutional nitrogen atom and adjacent vacancy aligned along a $\langle$111$\rangle$ crystallographic axis (Fig.\,\ref{fig:principle}e). An additional captured electron gives the defect a negative charge, and a deep potential well surrounding the vacancy tightly binds the valence electrons \cite{hossain08} enabling production of centres in nanocrystals less than 10\,nm in size \cite{tisler09, bradac10}. Far-field optical readout of the spin-1 ground state at room-temperature \cite{gruber97} places NV at a distinct advantage over other solid state qubits as a non-invasive sensor, since remote access and control is possible without (electrical) contact. The NV system shows great promise for nanoscale magnetometry as evident from recent demonstrations for static (DC) and oscillating (AC) fields \cite{bala08,maze08Nature,waldherr12,nusran12,maletinsky12}, wide-field imaging \cite{steinert10} and high sensitivity multi-pulse schemes \cite{hall10PRB,naydenovPRB11,delange11PRL}. Furthermore, the inertness of the diamond host crystal allows the NV system to be used unrestrictedly as a magnetic/fluorescent probe in biological environments \cite{mcguinness11}.\\

The decoherence of a coherent quantum system is determined by the noise distribution in the environment and interaction strength. For NV the Zeeman effect is larger than the Stark effect \cite{dolde11}, so it is magnetic fields rather than electric fields which produce the greatest contribution to decoherence. Local fields arise from macroscopic sources (such as the Earth's field), magnetic dipoles (from unpaired spins) and Biot-Savart fields (from moving charges). In ultra-pure environments, NV dephasing times exceeding a millisecond have been shown at room temperature \cite{bala09}, whereas for these nanodiamond studies, dipolar fields from internal nitrogen donors and spins on the diamond surface form intrinsic spin baths (see Fig.\,\ref{fig:theory}a) which restrict NV coherence times to below $10\,\mu$s \cite{tisler09}. Here we introduce spins external to the diamond lattice and surface, and detect their presence by measuring changes in the NV decoherence rate.\\

The effective Hamiltonian for an NV spin $\vec{S}$ in the presence of a magnetic field $\vec{B}(t)$ is:
\begin{equation}\label{eq:Ham}
\mathcal{H}_{\mathrm{NV}}=\hbar \vec{S}\mathbf{D}\vec{S}+\hbar \vec{S}\mathbf{A}\vec{I}+\hbar\gNV\vec{B}(t)\cdot\vec{S},
\end{equation}
where \gNV{} is the NV gyromagnetic ratio and $\mathbf{D}$, $\mathbf{A}$ are the fine and hyperfine tensors which produce splitting of the order $D \approx 3$\,GHz and $A \approx 2$\,MHz (for $^{14}$N with nuclear spin $I=1$) and act to partially lift the degeneracy of the NV ground state (see Fig.\,\ref{fig:principle}e).\\

We define the fluctuation regime of the magnetic environment in which the sensor is placed by the dimensionless parameter $\Theta=\fe/ \gNV \Bsd{e}$ where \fe{} is the average fluctuation rate of the field and \Bsd{e} its fluctuation amplitude (different environments will be denoted with a subscript) \cite{hall09}. The random processes we are concerned with are characterized by zero mean fields, so that \Bsd{e} $\approx$ \Bsd{RMS}. Rapidly (slowly) fluctuating fields are characterised by the $\Theta\gg1$ ($\Theta\ll1$) condition. We begin by describing the effect of the intrinsic NV environment without the addition of any target spins in solution.\\

Internal electronic spins from nitrogen dopants in the nanocrystals provide a spin bath in the $\Theta\sim \mathcal{O}(1)$ regime \cite{delange12}. Surface spin flips occur over faster timescales providing an additional spin bath in the $\Theta \gg1$ regime \cite{panich11}. In Fig.\,\ref{fig:theory}b) we plot the typical magnetic fields from these distinct sources resulting from a quantum cluster expansion calculation. The relative contribution of each spin bath can be observed by applying a spin-echo pulse sequence, which refocusses slow fluctuations but retains the effect of spin baths with relatively fast dynamics. Spin-echo is a common technique in magnetic resonance spectroscopy, although here it is applied to the NV spin, rather than to the sample. To describe the effect on the NV coherence quantitatively we use an overall function for the ground state probability after spin-echo time $t$ of the form:
\begin{equation}\label{eq:Total}
P(t)=\exp\left[-\GL{Ext}t-\frac{1}{\left(\GUL{short}{N}t\right)^{-4}+\left(\GUL{long}{N}t\right)^{-1}}\right],
\end{equation}
where $\GUL{short}{N}=f_{\mathrm{N}}/\sqrt{2\sqrt{2}\Theta_{\mathrm{N}}}$ and $\GUL{long}{N}=\gNV \Bsd{N}/2\Theta_{\mathrm{N}}$ are the decoherence rates from the internal nitrogen spin bath at short and long times \cite{hall09} and before the addition of external target \Mn{} spins the ``external'' decoherence rate ($\Theta \gg1$) is due to the surface spin bath
\begin{equation}\label{eq:surfaceG}
\GL{Ext}\rightarrow\GL{ss}=\frac{\gNV^2 B^2_{\mathrm{ss}}}{2f_{\mathrm{ss}}}.
\end{equation}
In Fig.\,\ref{fig:theory}c) we plot the spin-echo curves corresponding to the calculations of Fig.\,\ref{fig:theory}b) for different NV depth scenarios. When the NV is near the surface ($\sim$5nm) the spin-echo curve is dominated by the exponential curve of the fast fluctuating surface spins, whereas when the NV is further from the surface ($\sim$15nm) the spin-echo curve is also modulated at short times by the intrinsic nitrogen donor spin bath. Alternatively, the NV spin can be driven continuously with a microwave field to produce Rabi oscillations. The decay of the Rabi oscillations is also characteristic of the environmental noise spectrum \cite{dobrovitski09} however, in the regime dominated by surface spins we obtain essentially the same information as from the spin-echo measurement.\\

Upon introduction of \MnCl{} solution to the nanodiamonds, dipolar coupling to the additional magnetic sources ($S=5/2$ \Mn{} spins) alters the NV decoherence profile, providing new extrinsic decoherence channels (as opposed to the nitrogen and surface spin baths intrinsic to the nanodiamond) to modify Eq.\,\eqref{eq:Total} such that $\GL{Ext} = (\GL{ss} + \GL{Mn})$. Importantly, measurement of the spin-echo profile before the addition of spin solution allows the initial decoherence sources to be characterised. With knowledge of the initial spin environment, measurements on the same single spin probes allow us to detect changes in local magnetic fields arising from exposure to different solutions.\\

\section{\label{experiment}Detecting spins in solution using quantum decoherence}

\begin{figure*}
\includegraphics[width=\textwidth]{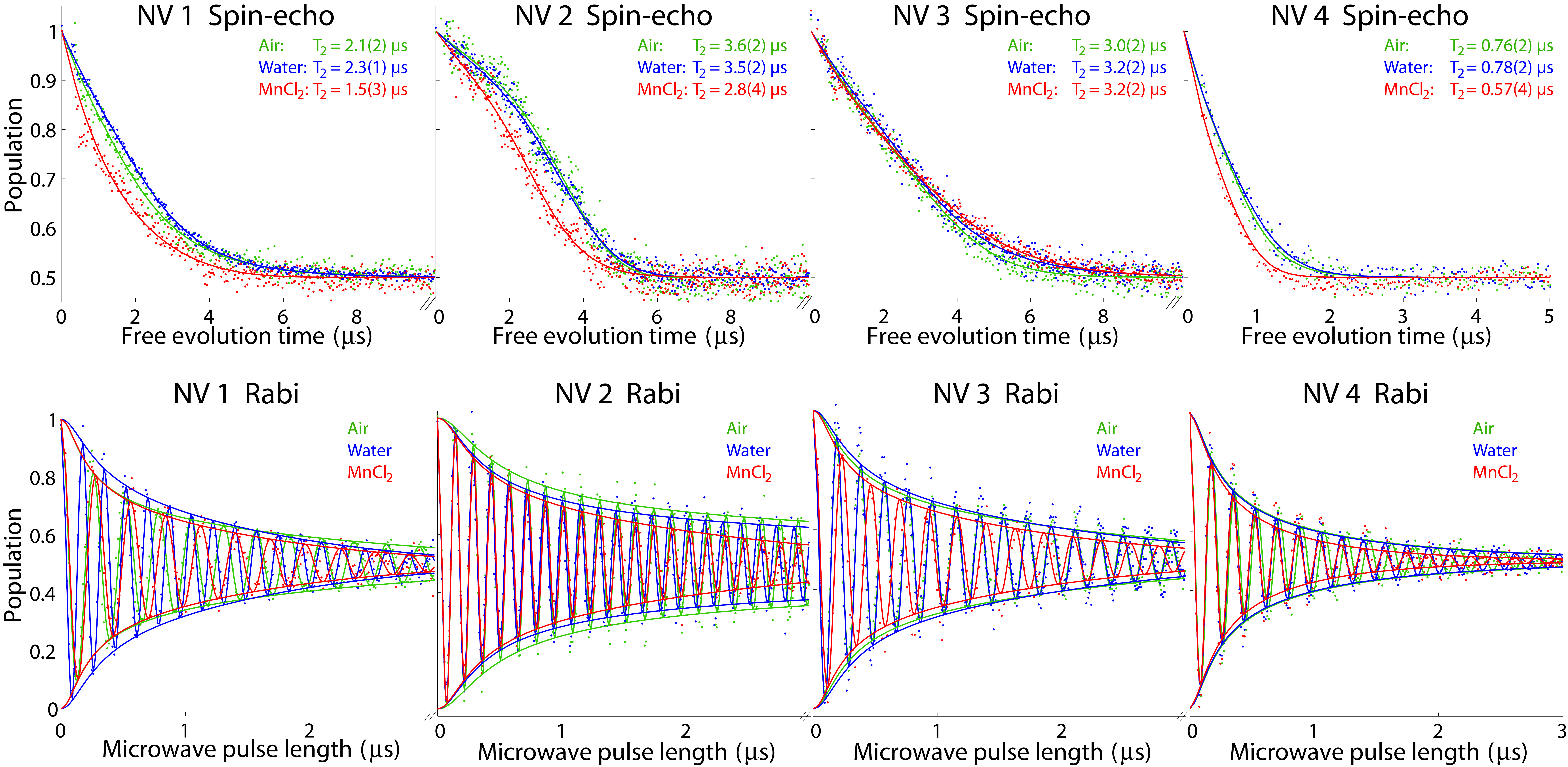}
\caption{Decoherence properties of NV centres under various immersion conditions and exposure to \Mn{} spins in solution. Top row: Measured spin-echo decoherence profiles for four NV centres, performed on nanocrystals in air (\colair), deionised water (\colwater), and 1M\,\MnCl{} (\colmn). Fits to the data-points using Eq.\,\eqref{eq:Total} (shown as the solid lines) allow the decoherence rates under the various immersion conditions to be determined (shown in the legend with uncertanties in the last significant figure given in parentheses). Bottom row: Rabi measurements corresponding to the spin-echo data for each NV.}
\label{fig:results}
\end{figure*}

Fig.\,\ref{fig:results} shows the spin-echo and Rabi profiles for four NV centres (referred to as NV\,1--4) in air and water (\colair{} and \colwater{} traces respectively). As expected the Rabi profiles correlate well with the spin-echo results. The diversity in decoherence rates, with coherence times ranging from 800\,ns to 3.5\,\us{}, indicate each NV experiences a unique nanoscale environment, due to variations in crystal size, purity and NV proximity to surface. Water is paramagnetic, but with a dipole moment three orders of magnitude smaller than the electron spin, hence the addition of water has little observable effect on the magnetic environment of the nanodiamonds as measured through \ttwo. Removal or deposition of impurities at the diamond surface in water, self diffusion of water molecules, and experimental uncertainties contribute to minor differences between the measured profiles.\\

The form of the spin-echo decay before the addition of the Mn spins can be used to estimate the contribution of each native spin bath to the overall decoherence. For example, NV\,2 exhibits Gaussian decay at short times characteristic of the internal nitrogen spin bath, whereas the exponential decay of NV\,4 indicates dominance of surface spins. In columns 1 and 2 of Table\,\ref{tab} we present the decoherence rate associated with the initial surface spin bath.\\

\begin{table}[h]
\caption{Decoherence rates $\GL{Ext}$ (kHz) extracted from the data in \autoref{fig:results} and \autoref{fig:HCl}. Fitting uncertainties in parentheses at the 95\% confidence level.\label{tab}}
\begin{ruledtabular}
\begin{tabular}{lcccc}
& \textbf{Air} & \textbf{Water} & \textbf{\MnCl} & \textbf{HCl}\\
NV1 & 410 (60) & 350 (15) & 660 (20) & 470 (20)\\
NV2 & 120 (20) & 140 (10) & 200 (40) & 90 (30)\\
NV3 & 270 (20) & 240 (30) & 280 (10) & -- \\
NV4 & 1,140 (100) & 1,130 (70) & 1,600 (70) & -- \\
\end{tabular}
\end{ruledtabular}
\end{table}

\subsection{Analysis of the initial surface spin bath}
We first consider the control measurements in air and water. In the absence of any \emph{a priori} knowledge of the initial environment, we describe the evolution of the NV and surface environment at a quantum level, using a master equation approach with a general relaxation rate $f_{\rm ss}$ due to the surface spin dynamics. To solve the system we use a cluster expansion method \cite{witzel05}, tracing over the bath degrees of freedom, to calculate the response of the NV centre to the surface and internal spin baths. From the results of the cluster expansion, we determine the following dependency of the magnetic field at the NV centre:
\begin{equation}\label{eq:surfaceB}
B_{\mathrm{ss}}=277 \,\mu \mathrm{T}\,\mathrm{nm}^3 \times \frac{\sigma^{1/2}}{h^2},
\end{equation}
where $\sigma$ is the effective surface spin density and \emph{h} is the depth of the NV below the diamond surface. Using Eq.\,\ref{eq:surfaceG}, this relationship allows us to compare the relative depths of each NV ($i=1,...,4$) below the surface, since $\Gamma^{(i)}_{\rm ss}/\Gamma^{(j)}_{\rm ss}=(h_j/h_i)^4$. For NV depths which are large compared to the spatial separation of surface spins, we can approximate the spin density and fluctuation rate of surface spins as being equivalent for all nanocrystals studied. Taking the depth of NV\,4 as a reference, we find $h_1=(1.3\pm0.03) h_4$; $h_2=(1.8\pm 0.02) h_4$; and $h_3=(1.4\pm 0.02) h_4$.\\

The control measurements  also provide insight into the mechanism of surface spin relaxation. Despite not knowing the precise physical origin of the surface spin relaxation, we may still rule out certain mechanisms as being responsible. Under a dipole-dipole mediated surface flip-flopping process, the cluster expansion analysis gives $f_{ss}=5.56\,\mathrm{GHz}\,\mathrm{nm}^3\times \sigma^{3/2}$. This expression, together with Eq.\,\ref{eq:surfaceG}, and the $\Theta\gg1$ condition allows us to set the following upper limit for the NV depth as a function of the decoherence rate
\begin{equation}\label{eq:height}
h\ll\left(\frac{2.28\,\mathrm{MHz}}{\GL{ss}}\right)^{1/3}\mathrm{nm}.
\end{equation}
From Eq.\,\ref{eq:height}, we see that even a decoherence rate as low as $\GL{ss} \simeq$\,100\,kHz implies an NV depth much less than 2\,nm below the nanocrystal surface, below the known photo-stability limit \cite{bradac10}. In contrast, spin-orbit coupling of sp$^2$ hybridised surface carbon to phonons results in surface spin relaxation rates of 0.1 -- 10\,GHz at 300\,K \cite{panich11} independent of the spin density and is therefore the more likely mechanism.

\subsection{Detection of extrinsic manganese spins}
In Fig.\,\ref{fig:results} (\colmn{} traces) we show the effects of immersion of the nanodiamonds in 1\,M \MnCl{} (in HCl) on the spin echo decay envelopes. \Mn{} ions were chosen due to their high spin ground state [Ar]3d$^5$ at room temperature, characterised by an electronic spin of S=5/2. A strong acid was used to prevent oxidation of \Mn{} to a low spin state. In addition to a reduction in the coherence time, the spin-echo profiles of NVs\,1, 2, and 4 become noticeably more exponential, characteristic of an additional fast magnetic noise contributing to the total profile $P(t)$. In contrast, NV\,3 experiences no change within the bounds of uncertainties, despite having initial decoherence comparable to the three other NV centres- implying a similar proximity to the surface. The absence of any change is attributed to close proximity to a surface not exposed to the \MnCl{} solution (e.g. the coverslip, or another nanodiamond). We observed that a number of other NV centres experienced minimal changes when immersed in manganese solution, which is also consistent with a greater physical separation between the NV and target manganese spins, e.g. $h > 10$ nm in the intrinsic nitrogen bath regime (Fig.\,\ref{fig:theory}c)). AFM profilometry shows a maximal NV distance from the external Mn spins of ~50 nm, further highlighting the nanoscale nature of the detection.\\

\begin{figure}
\includegraphics[width=12cm]{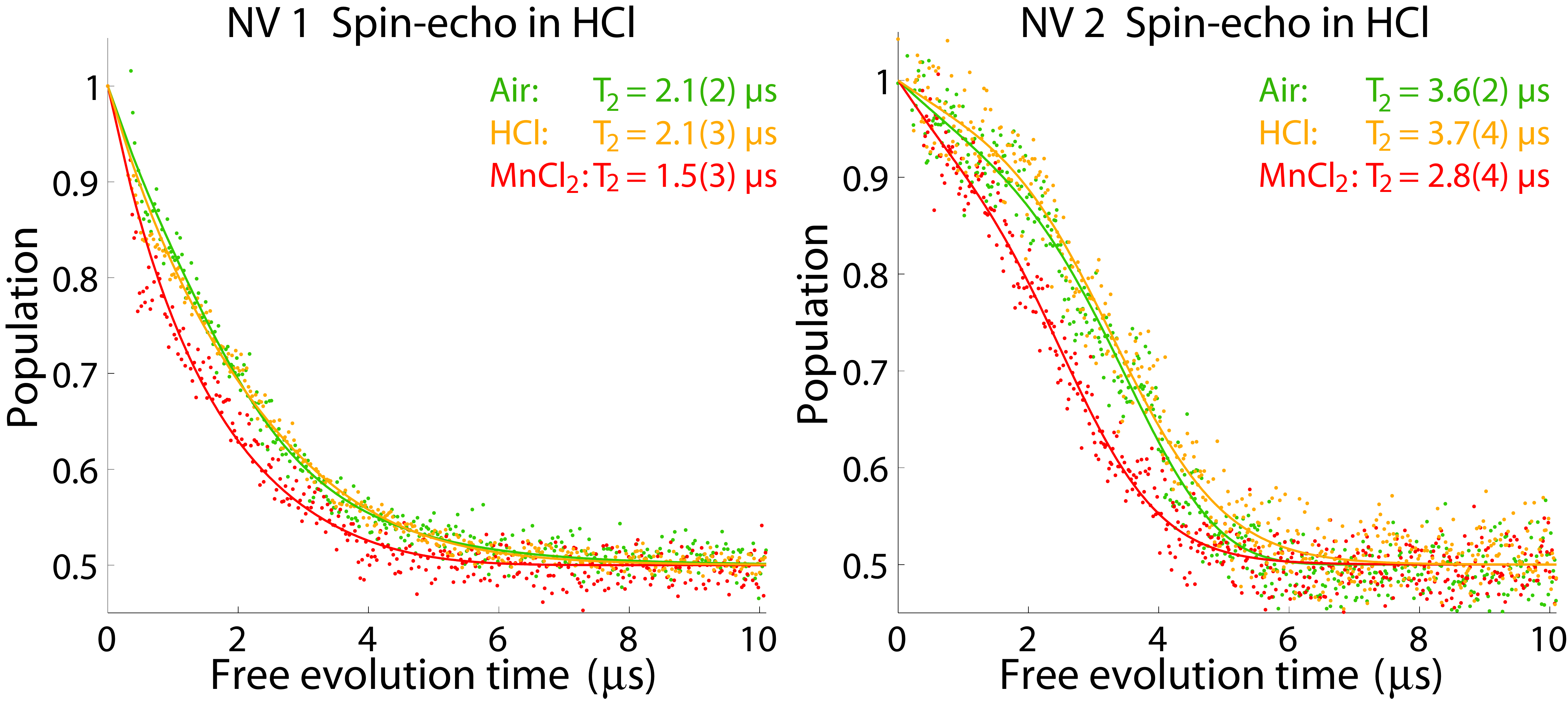}
\caption{Restoration of NV quantum coherence after washing in spin-free acid. Measured spin-echo decoherence profiles for NV\,1 and 2, performed in air (\colair), 1M \MnCl{} (\colmn), 1M HCl (\colhcl). Fits to the data-points using Eq.\,\eqref{eq:Total} are shown as the solid lines (fit uncertanties in the last significant figure given in parentheses). The same spin-echo profiles in air and HCl confirm Mn spins are responsible for the observed changes in decoherence.}
\label{fig:HCl}
\end{figure}

As a negative control, we measured the decoherence of NV1 and NV2 when immersed in a 1\,M HCl solution, possessing  an equivalent concentration of ions to the 1\,M \MnCl{} solution previously detected. The restoration of quantum coherence (Fig.\,\ref{fig:HCl}) confirms that the $T_2$ based decoherence detection is primarily sensitive to the presence of \Mn{} spins rather than changes in pH, ionic diffusion, or charge transfer. The measured rates in HCl, also allow us to eliminate ambient contributions to the decoherence rate, and isolate the contribution from the \Mn{} spins alone.\\

We may gain insight into the dynamics of the manganese spins using a scaling analysis to determine the dependence of the fluctuation rate on the NV depth. No depth dependence would imply an effectively frozen lattice type dipolar interaction between the \Mn spins \cite{khutsishvili69, wrachtrup95, desousa05}, whereas a $f_\mathrm{Mn}\propto h^{-2}$ dependence would imply a pure diffusion process with frozen magnetisation. As such, we set $f_\mathrm{Mn} = \alpha h^x$ and proceed as follows to determine the scaling exponent, $x$, from the data. We first express the ratios of decoherence rates due to Mn in NV1 and NV2 as
\begin{eqnarray}
 \frac{\Gamma^{(2)}_\mathrm{Mn}}{\Gamma^{(1)}_\mathrm{Mn}} =    \frac{\left(B^{(2)}_\mathrm{Mn}\right)^2f^{(1)}_\mathrm{Mn}}{\left(B^{(1)}_\mathrm{Mn}\right)^2f^{(2)}_\mathrm{Mn}} = \left(\frac{h_1}{h_2}\right)^{x+4}
\end{eqnarray}From Eq.\,\ref{eq:surfaceG}, we had $\frac{h_1}{h_2} = \left(\frac{\Gamma^{(2)}_\mathrm{ss}}{\Gamma^{(1)}_\mathrm{ss}}\right)^{1/4}$, and taking $\Gamma_{{\rm Mn}} \sim \Gamma_{{\rm Ext}}[{\rm MnCl}_2] - \Gamma_{{\rm Ext}}[{\rm HCl}]$
%
%
we have
\begin{eqnarray}
  x &=& 4\ln\left(\frac{\Gamma^{(2)}_\mathrm{Ext}[{\rm MnCl}_2] - \Gamma^{(2)}_\mathrm{Ext}[\mathrm{HCL}]}
  {\Gamma^{(1)}_\mathrm{Ext}[{\rm MnCl}_2] - \Gamma^{(1)}_\mathrm{Ext}[\mathrm{HCL}]}\right) /\ln\left(\frac{\Gamma^{(2)}_\mathrm{ss}}{\Gamma^{(1)}_\mathrm{ss}}\right)-4\nonumber\\ &=& -2.2 \pm 1.2,
\end{eqnarray}which suggests that the majority of the decoherence may be attributable to the self-diffusive motion of the Mn spins. The dominance of diffusion in the NV dephasing over direct Mn-Mn flip-flopping processes is perhaps not unexpected since the motion of the Mn spins reduces the mean time for dipole interaction between the Mn spins (motional narrowing). From an analysis of the magnetic field generated by motionally diffusing Mn spins in 1M \MnCl{} concentration we determine an upper bound on the number of spins detected at a maximal detection standoff of 10 nm (beyond which the NV is dominated by the intrinsic spin-bath -- see cluster expansion calculations in Fig.\,\ref{fig:theory}c). Using NV4 as an example, the data corresponds to an upper bound detection of 2,500 Mn spins, over an effective solution volume of (16 nm)$^3$ in a single-time-point acquisition time of 4.2 seconds. This is an improvement of several orders of magnitude in resolution and sensitivity over state-of-the-art ESR detection of $2\times 10^7$ spins over micron scales  in 3600 s \cite{blank04}. Due to the $h^{-3}$ scaling of the external field variance giving rise to the decoherence change signal, reducing the size of the nanodiamond will dramatically improve the sensitivity to small spin ensembles: e.g. for NV in 7 nm nanodiamond (near the photostability limit) the results indicate that less than 100 spins could be detected, with a time resolution down to 400 ms.\\

\begin{figure}
\includegraphics[width=12cm]{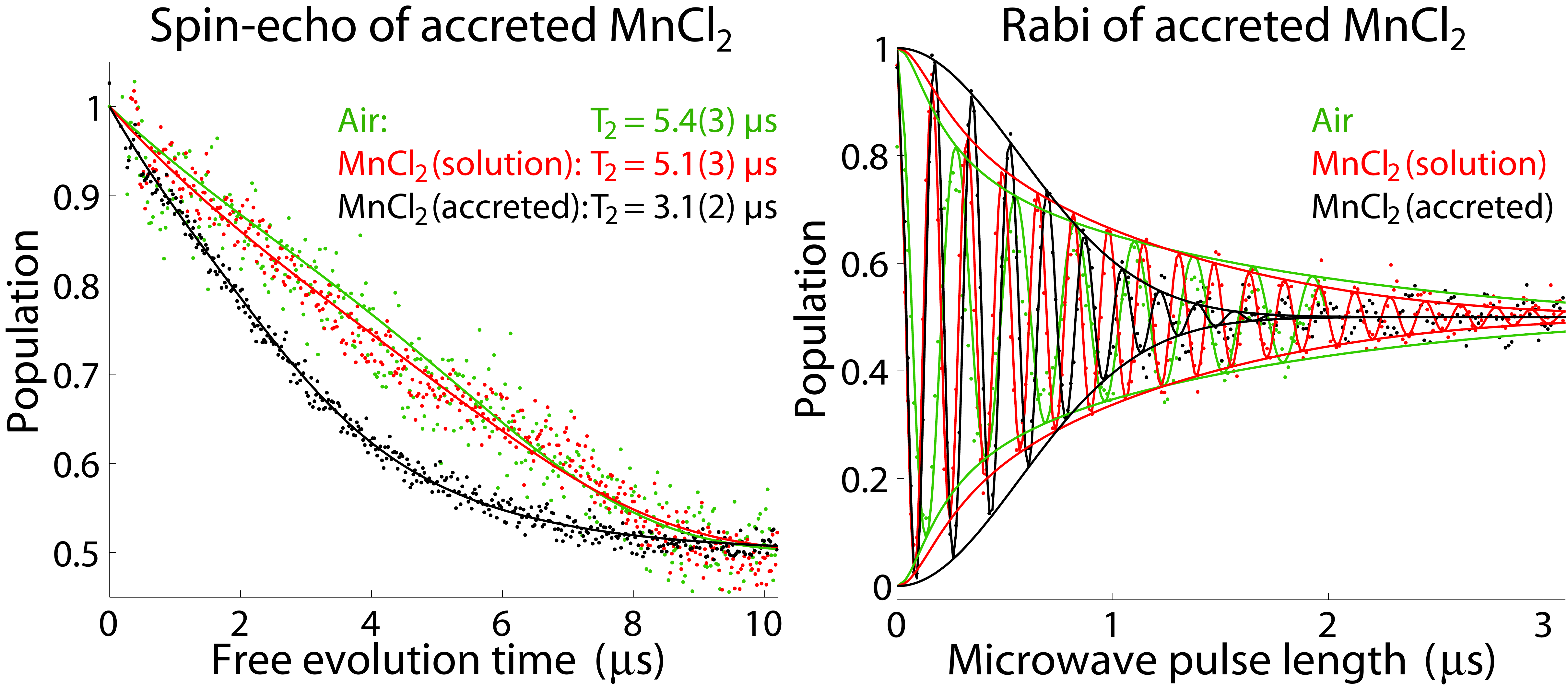}
\caption{Measurements on NV5 showing detection of \Mn{} spin accretion on the nanodiamond surface. Rabi and spin-echo measurements in air (\colair), 1M \MnCl{} solution (\colmn), accreted \MnCl{} (\colacc) (fit uncertanties in the last significant figure given in parentheses).}
\label{fig:accretion}
\end{figure}

The decoherence sensing technique could equally be used to detect changes in spin density on the nanodiamond surface; for instance in monitoring chemical reactions at the surface of functionalised nanodiamonds. In a final experiment we characterised another NV centre (NV\,5), which was observed to have decoherence rates of $\Gamma_{\mathrm{Ext}}[{\rm Air}] = 136$\,kHz and $\Gamma_{\mathrm{Ext}}[{\rm MnCl}_2] = 160$\,kHz. In this case the solution was left to evaporate in air resulting in precipitation from the \MnCl{} solution to the nanodiamond surface. Upon re-suspension in deionised water, there was a significant increase in decoherence rate to 260\,kHz, as can be seen in Fig.\,\ref{fig:accretion} (\colacc{} trace). We attribute this to an increased surface spin density due to additional \Mn{} spins on the nanodiamond surface.

\vspace{-3mm}
\section{Conclusion and outlook}
We have demonstrated, for the first time, nanoscale magnetic detection of spins extrinsic to the sensor medium, using decoherence of single NV spins. From the NV spin-echo and Rabi profiles we observed changes in decoherence rates associated with diffusion and accretion processes of atomic spins in solution. Our measurements over several NV defects in separate nanodiamonds provide information on the relative location of the NV centres with respect the diamond surface and dynamic properties of the external spin environment. Our results also shed new light on the surface spin dynamics of nanodiamond. The data is supported by a detailed cluster expansion treatment and indicate the detection of order $10^3$ spins with nanoscale resolution in a single-time-point resolution of 4.2 s, which is a significant improvement over current high-resolution detection of electronic spins under ambient conditions \cite{blank04}. The NV-decoherence based sensing technique is a new and promising approach to ambient nanoscale detection of magnetic field fluctuations, breaking through the fundamental mesoscopic spatial limits imposed by conventional MR techniques. Detection using single NV spin qubit probes may be applied to arbitrary spin samples including solid-state, chemical, or biological environments where no other sensor currently exists with the desired combination of sensitivity and resolution. Our results correspond to \ttwo-based detection, however similar measurements using a \tone-based scheme \cite{cole09} will also open a new regime of fluctuation detection. For schemes where ultimate spatial resolution is not required, the use of diamonds with high density NV ensembles will allow $\sqrt{N}$ improvement in sensitivity. By reducing the size of the nanodiamonds to 7\,nm \cite{bradac10}, the sensitivity of the NV centre to its environment will enable single point detection times well below 1 s and detection of much smaller target ensembles, opening a new vista of novel nanomagnetic investigations of biological processes. Recent experimental demonstrations of the quantum control techniques required for decoherence detection in the complex environment of living cells \cite{mcguinness11}, underpin the considerable potential of nanoscale decoherence-based biosensing \cite{hall10PNAS}.\\

\vspace{-5mm}
\section{Acknowledgements}

The authors thank S. Steinert for a critical reading of the manuscript. This work was supported by the Australian Research Council under the Centre of Excellence (CE110001027) and Discovery Project (DP0770715) schemes, ERC Project SQUTEC, EU Project DINAMO, and the Baden-Wuerttemberg Foundation.

\end{document}